\documentstyle[aps,prb,twocolumn,epsfig]{revtex}

\begin{document}
\pagestyle{myheadings}

\wideabs{
\draft
\title{Raman Scattering from Magnetic Excitations in coupled Spin Ladders}
\author{R. Citro$^{a}$ and E. Orignac$^{b}$}
\address{
$^{a}$Dipartimento di Scienze Fisiche ''E.R. Caianiello'',\\
University of Salerno and Unit\`{a} INFM of Salerno, Baronissi (Sa), Italy\\
$^{b}$ Laboratoire de Physique Th\'eorique, CNRS UMR 8549, \\
 Ecole Normale   Sup\'erieure, 24 Rue Lhomond 75231 Paris Cedex 05, France \\}
\date{\today}
\maketitle\thispagestyle{myheadings}

\begin{abstract}
We consider Raman scattering of coupled two leg
spin ladders in the Fleury-Loudon regime. We derive the dependence of
the intensity with 
polarization of the incoming light and temperature and discuss the
effect of interaction of the elementary excitations on the shape of
the spectrum. We show that
Raman scattering spectrum is sensitive to the effective dimensionality
of the coupled ladder system, making Raman scattering useful to
discriminate true ladder systems from quasi two dimensional spin gap
systems.  
\end{abstract}

\pacs{ PACS numbers: 72.12.Di 87.64.Je 75.10.Jm}
}

%\narrowtext
The study of spin ladder compounds~\cite{dagotto_2ch_review} has become
a subject of intense theoretical and experimental
activity\cite{johnston_ladder_suscep} in the recent years. Insulating
two leg spin-ladder compounds present a spin gap state that has been
probed experimentally by thermodynamic (specific heat and magnetic
susceptibility) measurements as well as dynamical measurements such as
neutron scattering or NMR.  
Another technique to study the dynamics of magnetic excitations in low
dimensional antiferromagnets is  Raman scattering. This technique  has
been used in particular to probe spin
1/2 chains \cite{yamada_raman_1d}, spin 1 chains \cite{sulewski_raman_spin1}
and spin Peierls systems
\cite{gugeo3_lemmens,loosdrecht_raman_cugeo3,kuroe_raman_nav2o5,golubchik_raman_nav2o5}.
The extended interest towards this experimental technique is due
 to its sensitiveness to singlet excitations, that render it
complementary to neutron diffraction experiments.  
There exists at present a certain
amount of literature on the theory of Raman scattering from dimerized spin chains,
both analytical \cite{uhrig_dimerized,brenig_raman_frustrated} and numerical
\cite{augier} in relation to experiments on Spin-Peierls
compounds\cite{gugeo3_lemmens,loosdrecht_raman_cugeo3,kuroe_raman_nav2o5,golubchik_raman_nav2o5}.  
In the case of spin ladder materials, although there are quite a few
experimental
investigations\cite{srcuo_raman_spectra,cavo_raman_spectra,sugai_raman_lacuo25,sugai_raman_sr14cu24o41,popovic_raman_sr14cu24o41} 
by Raman scattering,
there is a scarcity of theoretical results on this subject. While some numerical calculations are available to describe the experimental data
\cite{spinladder_raman_numerics}, no analytical expressions of the Raman intensity for
$2n$-legged spin-1/2 ladders existed. In all
the theoretical work on spin ladders, the analysis
of magnetic Raman scattering is based on the Fleury-Loudon
Hamiltonian\cite{fleury_raman,moriya_raman,shastry_raman}, 
that describes the interaction of photons with
magnetic excitations. This limits the investigations to the non-resonant
regime\cite{shastry_raman}. In a previous report~\cite{orignac_raman_short},
we have derived the relevant expressions for two leg ladders  with weak
 and strong interladder coupling. In our report~\cite{orignac_raman_short}, we found that for
weak coupling case the 
Majorana fermions description of the spin
ladder~\cite{shelton_spin_ladders}  lead to a cusp in the Raman
intensity at twice the gap. This was in disagreement with
experiments that show a peak rather than a cusp in the Raman intensity
for a frequency twice the gap. On the other hand, in the strong
coupling case,  the Bond Operator
Technique~\cite{chubukov89,sachdev_bot,gopalan_2ch} (BOT) 
 predicted correctly the presence of the peak at twice the gap. 
This suggests that the strong coupling description is
more adequate to describe Raman scattering experiments in 
real spin ladder systems.
In our report, some important aspects of Raman scattering in spin
ladder were not considered. First, we restricted ourselves to zero
temperature. However, the spin gap is temperature dependent and thus
temperature should modify the Raman intensity. A second aspect that
was not considered is the effect of interactions between the magnetic
excitations. Such interactions break the coherence that gives rise to
the peaks and it is important to understand how they modify the ideal
spectrum of Ref.~\onlinecite{orignac_raman_short}. A last aspect is
the influence of interladder excitations. There are spin ladder
materials\cite{cavadini_kcucl3_neutrons,garrett_vo2p2o7} 
in which the interladder coupling is by no means negligible
such as $\mathrm{KCuCl_3}$ or $\mathrm{VO_2P_2O_7}$. This affects
strongly the neutron scattering intensity compared to the one of a
single ladder. This two-dimensionality should also affect the Raman
spectrum. Understanding how the  Raman intensity is affected by
dimensionality  is  important to discriminate a one-dimensional
from a two-dimensional compound. A possible application could be to the
CuHpCl compound for which there exists some evidence of two
dimensional behavior\cite{hammar_transition_cuhpcl} (See however
Refs.~\onlinecite{chaboussant_cuhpcl,chaboussant_nmr_diaza,chaboussant_mapping}).   

In the present paper we deal with the calculation of the Raman scattering
cross section of  coupled spin ladders. After recalling
some basic results on the Fleury-Loudon theory of magnetic Raman scattering,
and the Bond Operator Technique\cite{sachdev_bot,gopalan_2ch}
(BOT)  we give the general derivation of the Raman
operator for coupled ladders in Sec. \ref{sec:loudon_fleury}. In Sec.
\ref{sec:isolated_ladder} we give an expanded discussion of the two leg
Heisenberg ladder\cite{orignac_raman_short}.  We discuss the 
 temperature dependence of the Raman intensity of the ladder. We also
consider the effect of interactions on the peaks in the Raman
spectrum and derive a scaling form of the intensity valid in the
vicinity of the edges of the Raman spectrum. 
Having illustrated the technique on and the physics on the simple
isolated ladder case, we turn   in Sec. \ref{sec:two_dimensional}
to the case of coupled ladders. First, we consider a system of two
coupled ladders, for which the Raman spectrum presents four peaks. The
influence of temperature and interactions on the spectrum is the same
as in the case of the isolated ladder. Then, we consider the case of a
two dimensional array of ladders. We show that the spectrum of this
system is markedly different from that of the isolated ladder, and
discuss the effect of the interactions of magnetic excitations on the
singularities of the two-dimensional spectrum. 

\section{The Loudon-Fleury theory for the periodic array of
ladders}\label{sec:loudon_fleury}

We consider a periodic array of single-rung two-legged Heisenberg ladders
(shown in Fig.\ref{fig:array}) whose Hamiltonian is

\begin{eqnarray}\label{hladder_array}
H&= &\sum_{i,j}\left ( J_{\parallel}{\bf S}_{i,j}\cdot{\bf S}_{i+1,j}+
J_{\perp} {\bf S}_{i,2j}\cdot{\bf S}_{i,2j+1}\right. \nonumber \\ 
&+ & \left.
J_{\perp}' {\bf S}_{i,2j+1}\cdot{\bf S}_{i,2j+2} \right ),
\end{eqnarray}
\noindent where $J_{\parallel}=\lambda J_{\perp}$ is the
interaction along the legs of the ladder,  $J_{\perp}$ is the
rung interaction and $J_{\perp}'=
\lambda' J_{\perp}$ is the interladder coupling. The index $i$
and $j$ run along the leg and rung direction respectively. In the
limit $\lambda'=0$ the Hamiltonian reduces to a sum of independent
single-rung ladders, while in the limit
$\lambda=\lambda'=0$ (strong-coupling) the Hamiltonian
reduces to a sum over independent two spin rungs.
In this paper we are interested
in the regime  $\lambda'<\lambda<1$.

The interaction of light with
the antiferromagnetic fluctuations is described by Loudon-Fleury's \cite
{fleury_raman,moriya_raman} photon-induced super-exchange operator

\begin{equation}  \label{eq:loudon_fleury}
H_{R}=\sum_{{\bf l},{\bf l'}}\gamma_{{\bf l l'}}({\bf E_I}\cdot {\bf
\delta }_{{\bf l l'}})
({\bf E_S}\cdot {\bf \delta }_{{\bf l l'}})%
{\bf S}_{{\bf l}} \cdot {\bf S}_{{\bf l'}}  \label{hr}
\end{equation}
where ${\bf E_I}({\bf E _S)}$ are the incident (scattered) electric field
vectors of photon, and ${\bf \delta }_{{\bf l l'}}$ is a unit
vector connecting the lattice 
sites ${\bf l}$ and ${\bf l'}$, at which the spins ${\bf
S}_{{\bf l}}$ and
${\bf S}_{{\bf l'}}$ are located. 
A derivation of (\ref{hr}) starting from the Hubbard
Hamiltonian can be found in Ref. \onlinecite{shastry_raman}.

The Raman cross section\cite{loudon_raman,reiter_raman}  can be expressed in terms of
of the retarded Raman response function as:
\begin{equation}
\frac{d^{2}\sigma }{d\Omega d\omega _{2}}=\frac{\omega_1 \omega_2^3}{2\pi
c^4 V} \frac {n_2}{n_1} \frac 1 {1-e^{-\beta \hbar \omega}} {\rm Im}
\chi_R(\omega)
\end{equation}
$\omega _{1}$ and $\omega _{2}$ are the frequencies of the incoming and
scattered radiation, respectively, $\omega=\omega_2-\omega_1$, $n_1$
and 
$n_2$
are the refractive indices of the material at frequency respectively
$\omega_1$ and $\omega_2$. $V$ is the volume of the crystal and $c$
the velocity of light.  The retarded linear response function $\chi_R(\omega)$ is defined as:
\begin{equation}
\chi _{R}^{ret}(\omega)=\frac{\imath}{\hbar}\int_0^{\infty}
e^{\imath(\omega+\imath 0) t} {\rm Tr}\left\{ Z^{-1}e^{-\beta H}\left[
H_{R}(t),H_{R}(0)\right] \right\} ,  \label{ramansusc}
\end{equation}
where $Z={\rm Tr}e^{-\beta H}$and $H_{R}$ is the Loudon-Fleury Hamiltonian
 (%
\ref{eq:loudon_fleury}).

By inserting the resolution of identity in (\ref{ramansusc}), the Raman
intensity can be written as

\begin{equation}
\frac{d^{2}\sigma }{d\Omega d\omega _{2}} \propto \frac{1}\hbar\frac 1 Z
\sum\limits_{n,m}e^{-\beta E_{n}}\left| \left\langle \Psi _{n}\left|
H_{R}\right| \Psi _{m}\right\rangle \right| ^{2}\delta (\omega
-(E_{n}-E_{m})/\hbar),  \label{ramanme}
\end{equation}
where $\mid \Psi _{n(m)}\rangle $ are eigenstates with energies $E_{n(m)}.$
Such formula can be easily interpreted as a Fermi golden rule averaged over
the Boltzmann weight. To obtain informations on two-magnons scattering
processes we should perform a symmetry analysis of the matrix elements
appearing in (\ref{ramanme}), and discuss selection rules. Since the spin
ladder Hamiltonian is invariant under translation along the legs, SU(2)
rotation, and mirror along the  the leg direction, an eigenstate should be
characterized by a (lattice) momentum defined modulo $2\pi /a$ (where $a$ is
the lattice spacing), a spin and its parity under leg exchange. The Raman
operator defined in (\ref{hr}) is rotationally and translationally
invariant, and still invariant under leg exchange. As a result, the
selection rules impose that the states $\mid \Psi _{n}\rangle $ and $\mid
\Psi _{m}\rangle $ have the same spin, momentum and parity under leg
exchange. This implies in particular that at $T=0$, transitions will only
take place to states of total momentum zero, spin zero and same parity as
the ground state.

To obtain an expression of the Raman Hamiltonian more adapted to
perform the calculations, we introduce a coordinate system
$(\hat{x},\hat{y})$ where $\hat{x}$ is parallel to the chain direction
and $\hat{y}$ is parallel to the rung direction. In this coordinate
system, we have ${\bf l}=i\hat{x}+j\hat{y}=(i,j)$, 
${\bf E}_I=E_I(\cos \theta_I \hat{x} + \sin \theta_I
\hat{y})$ and ${\bf E}_S=E_S(\cos \theta_S \hat{x} + \sin \theta_S
\hat{y})$. We define $\gamma_\parallel=\gamma_{(i,j),(i+1,j)}$,
$\gamma_\perp=\gamma_{(i,2j),(i,2j+1)}$ and
$\gamma'_\perp=\gamma_{(i,2j+1),(i,2j+2)}$. 
The expression of the Raman operator becomes:
\begin{eqnarray}
H_R &= &E_I E_S \cos \theta_S \cos \theta_I \gamma_\parallel \sum_{i,j}
{\bf S}_{i,j}\cdot{\bf S}_{i+1,j}   + E_I E_S \sin \theta_I \sin
\theta_S \nonumber \\ &\times & \left( \gamma_\perp \sum_{i,j} {\bf S}_{i,2j}\cdot{\bf
S}_{i,2j+1} + \gamma'_{\perp}\sum_{i,j} {\bf S}_{i,2j+1} \cdot  {\bf
S}_{i,2j+2} \right)
\end{eqnarray}

This expression can be simplified using the form of the full
Hamiltonian.
One has:
\begin{eqnarray}\label{eq:raman_op_array}
H_R = \frac{\gamma'_\perp}{J'_\perp} E_I E_S \sin \theta_I \sin
\theta_S H \nonumber \\ 
+ E_I E_S \left(\gamma_\parallel \cos \theta_S \cos
\theta_I - \frac{\gamma'_\perp J_\parallel }{J'_\perp} \sin \theta_I
\sin \theta_S \right) \sum_{i,j}
{\bf S}_{i,j}\cdot{\bf S}_{i+1,j} \nonumber \\
+  E_I E_S \sin \theta_I \sin
\theta_S\left(\gamma_\perp -\frac{\gamma'_\perp J_\perp}{J'_\perp}\right) \sum_{i,j} {\bf S}_{i,2j}\cdot{\bf
S}_{i,2j+1}
\label{hrgeneral}
\end{eqnarray}
The part of $H_R$ proportional to $H$ gives no contribution to the
response functions.

In the case of a single ladder, $J'_\perp=0$ and the simplified form
of $H_R$ is:
\begin{eqnarray}\label{eq:raman_op_sl}
H_R^{sl}& = &\left(\gamma_\parallel \cos \theta_I \cos
\theta_S - \frac{\gamma_\perp J_\parallel}{J_\perp}\sin \theta_I \sin
\theta_S \right) \nonumber \\ &\times & E_I E_S \sum_{i,j}  \sum_{i,j}
{\bf S}_{i,j}\cdot{\bf S}_{i+1,j}
\end{eqnarray}
Some remarks on the coefficients $\gamma_\alpha$,
($\alpha=\parallel,\perp$) are in order here. According to
perturbative calculations\cite{shastry_raman} for a half filled
Hubbard ladder, one should have
\begin{equation}\label{eq:perturbative_condition}
\gamma_\perp/J_\perp=\gamma'_\perp/J'_\perp=\gamma_\parallel/J_\parallel. 
\end{equation}
With this relation, (\ref{eq:raman_op_sl}) is recovered when
taking the limit $J'_\perp \to 0$ in (\ref{eq:raman_op_array}).
In such case, the simplified form of the Raman Hamiltonian is:
\begin{eqnarray}
H_R=\gamma_\parallel E_I E_S \cos (\theta_I + \theta_S) \sum_{i,j}
{\bf S}_{i,j}\cdot{\bf S}_{i+1,j}
\end{eqnarray}
This form has been previously obtained in Ref. \onlinecite{freitas_raman}. It leads
to the same Raman intensity for $\theta_I=\theta_S=0$ and
$\theta_I=\theta_S=\pi/2$ in disagreement with experimental
data\cite{freitas_raman}. It was shown that treating the $\gamma$'s as
phenomenological coefficients and assuming $\gamma_\alpha
\propto \sqrt{J_\alpha}$ lead to better agreement with
experiments\cite{freitas_raman}. In the following, we will treat the
$\gamma$'s as phenomenological coefficients and then specialize to the
simplified expression that result when
(\ref{eq:perturbative_condition}) is assumed.  

To analyze the Raman susceptibility, we use
the Bond Operator Representation (BOT)~\cite{sachdev_bot,chubukov89} 
of quantum S=1/2 spins
used by Gopalan, Rice and Sigrist\cite{gopalan_2ch} in their mean field
approach to spin ladders. In this representation, one starts from weakly
coupled rungs, i.e. $\lambda=0$, and introduces on each rung a singlet $s^\dagger$ and three
triplets $t^\dagger_\alpha$($\alpha=x,y,z$) boson creation operators, that
span the Hilbert space of a single rung when acting on a vacuum state. Since
the rung can be in either the singlet or one of the triplet states, the
condition :
\begin{equation}  \label{eq:projection}
s^\dagger s + \sum_\alpha t^\dagger_\alpha t_\alpha=1
\end{equation}
has to be satisfied by the physical states. The representation of the spins
 $%
{\bf S}_{i,2j}$ and ${\bf S}_{i,2j+1}$ in terms of these singlet and triplet
operators, is derived in Ref.\cite{sachdev_bot,gopalan_2ch}. It reads:
\begin{eqnarray}
S_{i,2j}^\alpha=\frac 1 2 (s^\dagger t_\alpha + t^\dagger_\alpha s-i
\epsilon_{\alpha \beta \gamma} t^\dagger_\beta t_\gamma)_{i,2j} \\
S_{i,2j+1}^\alpha=\frac 1 2 (-s^\dagger t_\alpha - t^\dagger_\alpha s-i
\epsilon_{\alpha \beta \gamma} t^\dagger_\beta t_\gamma)_{i,2j}
\label{repr}
\end{eqnarray}
The Hamiltonian Eq.~(\ref{hladder_array}) can be rewritten in the BOT
as: 
\begin{eqnarray}
H &=& \left( \frac{J_{\perp }}{4}-\mu \right) \sum_{i,j,\alpha }t_{i,j,\alpha
}^{\dagger }t_{i,j,\alpha }\nonumber \\ & + &\frac{\lambda J_{\perp}s^{2}}{2}\sum_{i,j,\alpha
}(t_{i,j,\alpha }+t_{i,j,\alpha }^{\dagger })(t_{i+1,j,\alpha }+t_{i+1,j,\alpha
}^{\dagger }) \nonumber \\ \label{hamda} &-&\lambda ^{\prime }J_{\perp
}\frac{s^{2}}{4}\sum_{i,j=1,2}(t_{i,j,\alpha }+t_{i,j,\alpha }^{\dagger
})(t_{i,j+1,\alpha }+t_{i,j+1,\alpha }^{\dagger })\nonumber \\ 
& + &\sum_{i,j}\left (
-\frac{3}{4}J_{\perp }s^{2}-\mu s^{2}+\mu \right ) + \text{quartic terms}.
\end{eqnarray}
The quartic terms in $t,t^\dagger$ in Eq.~(\ref{hamda}) describe
interactions between triplet excitations. Their explicit form will not
be needed here.

Substituting
the operator representation of spins Eq.~(\ref{repr}) into the Raman
operator for 
coupled ladders Eq.~(\ref{hrgeneral}), one ends up with the following
expression: 

\begin{eqnarray}
H_R &= &E_I E_S \left(\gamma_\parallel \cos \theta_S \cos
\theta_I - \frac{\gamma'_\perp J_\parallel }{J'_\perp} \sin \theta_I
\sin \theta_s \right)\nonumber \\ 
&\times & \frac{s^2}{4} \sum_{i,j,\alpha}
(t^\dagger_{i+1,2j,\alpha}+t_{i+1,2j,\alpha})(t^\dagger_{i,2j,\alpha}+t_{i,2j,\alpha})
\nonumber \\
& + & E_I E_S \sin \theta_I \sin
\theta_S\left(\gamma_\perp -\frac{\gamma'_\perp
J_\perp}{J'_\perp}\right)  \sum_{i,j,\alpha} \frac 1 4
t^\dagger_{i,2j,\alpha} t_{i,2j,\alpha}
\end{eqnarray}

In Fourier space, this expression becomes:
\begin{eqnarray}
H_R = \frac{E_I E_S}{4} \sum_{{\bf k},\alpha} \left ( A_{\bf k}  t^\dagger_{{\bf k},\alpha} t_{{\bf
k},\alpha} + B_{\bf k} (e^{i k_x a} t^\dagger_{{\bf
k},\alpha}t^\dagger_{-{\bf k},\alpha} + \text{H. c.})\right )
\label{generalr}
\end{eqnarray}
where:
\begin{eqnarray}
A_{\bf k}&= &(\gamma_\perp-\frac{\gamma'_\perp J_\perp}{J'_\perp})
\sin \theta_I \sin \theta_S \nonumber \\ 
&+& 2s^2\cos (k_x a)(
\gamma_\parallel \cos \theta_I \cos \theta_S -
\frac{\gamma'_\perp J_\parallel}{J'_\perp} \sin \theta_I \sin \theta_S) \nonumber \\
B_{\bf k}&=& s^2\gamma_\parallel \cos \theta_I \cos \theta_S
- s^2\frac{\gamma'_\perp J_\parallel}{J'_\perp} \sin \theta_I \sin \theta_S
\end{eqnarray}

In particular, if $\gamma_\parallel /J_\parallel =\gamma_\perp /J_\perp =\gamma'_\perp /J'_\perp$, we get the
following relation between the coefficients

\begin{equation}
\label{akvsbk}
A_{\bf k}=2B_{\bf k}\cos (k_x a)=2s^2\gamma_\parallel\cos k_xa\cos(\theta_I+\theta_S).
\end{equation}

The above expressions will be used in the following of the paper to predict
the Raman intensity.

\section{The single rung ladder}\label{sec:isolated_ladder}
\subsection{Mean field theory}
We start by considering a single ladder of
two strongly coupled antiferromagnetic $S=1/2$ Heisenberg chains, whose
Hamiltonian is

\begin{equation}
H=J_{\parallel}\sum_{i}\left( {\mathbf S}_{1,i}{\mathbf S}_{1,i+1}+%
{\mathbf S}_{2,i}{\mathbf S}_{2,i+1}\right) +J_{\perp }\sum_{i}%
{\mathbf S}_{1,i} \cdot {\mathbf S}_{2,i} \label{hladder}
\end{equation}
\noindent where $J_{\parallel}=\lambda J_{\perp}>0$ and $J_{\perp }>0$ denotes the intra- and inter- chain
antiferromagnetic interactions, respectively. 
Substituting the BOT operator representation of spins (\ref{repr})
into the original Hamiltonian, one
ends up with an Hamiltonian quartic in boson fields. Treating the singlet
operator in a mean field approximation and neglecting interactions among the
triplets, one obtains the following Hamiltonian quadratic in triplet
operators\cite{gopalan_2ch}:

\begin{eqnarray}
H_{MF}=(\frac{J_{\bot }}{4}-\mu )\sum_{i,\alpha }t_{i,\alpha }^{^{\dagger
}}t_{i,\alpha }\nonumber \\
+\frac{Js^{2}}{2}\sum_{i,\alpha }(t_{i,\alpha }^{^{\dagger
}}+t_{i,\alpha })(t_{i+1,\alpha }^{^{\dagger }}+t_{i+1,\alpha }).
\label{meanfieldh}
\end{eqnarray}

\noindent The chemical potential term $\mu$ guarantees that the
condition (\ref{eq:projection}) 
is satisfied on average. The parameters $\mu$ and $s$ are determined in a
self-consistent way by the 
minimization of the free-energy. The self consistent equations to be
solved are:

\begin{eqnarray}
\left (s^2-\frac{3}{2}\right ) &+& \int_0^\pi \frac{dk}{4\pi} \coth \left(\frac{\beta
\omega_k}{2}\right) \left[ \sqrt{1+d \cos k} \right. \nonumber \\
 &+ & \left. \frac{1}{\sqrt{1+d \cos k}}\right]=0\nonumber \\
\label{smu} 
\left ( \frac{3}{2}+2 \frac{\mu}{J_{\perp}}\right ) &+&
\frac{4\lambda} d \int \frac{dk}{4\pi} \coth \left(\frac{\beta \omega_k}{2}\right)
\left[ \frac{1}{\sqrt{1+d \cos k}}\right. \nonumber \\
 &-& \left. \sqrt{1+d \cos k}\right] = 0.
\end{eqnarray}
\noindent where
\begin{equation}
d=\frac{2 \lambda s^2}{(\frac{1}{4}-\frac{\mu}{J_{\perp}})}.
\end{equation}

After some simple manipulations, from equations (\ref{smu}) we obtain the following
equation for $d$

\begin{equation}
\label{eqd} d=\lambda \left[ 3-  \int_0^\pi \frac{dk}{\pi} \frac{\coth
\left(\frac{\beta \omega_k}{2}\right)} {\sqrt{1+d \cos k}}\right] .
\end{equation}

When this equation is combined with the first equation in (\ref{smu}), we can solve
them numerically to obtain $s^2$ and $d$. Only in the zero temperature limit these
equations decouple completely and reduce to the equations (2.19) of
Ref.\onlinecite{gopalan_2ch}. For the purpose of this paper, i.e. the analysis of the
Raman intensity, we discuss the temperature dependence of the singlet
order parameter $s(T)$ and $d(T)$. The numerical analysis shows (see Fig.({\ref{fig:sdvst})) that at high enough temperature
$s$ is very small and increases as temperature is lowered. 
Such behavior is expected since at high enough temperature bosons 
cannot condense. In this high temperature regime where $s(T)=0$ the
mean-field Hamiltonian (\ref{meanfieldh}) commutes with the Raman
operator, leading to
zero Raman intensity from magnetic scattering.
 For low enough temperature, $s(T)\ne 0$ and a Raman signal
appears. As temperature is lowered, the bandwidth of triplet
excitations increases and the width of the magnetic Raman scattering
increases. Let us point out that the mean field calculation of $s(T)$
is valid only for a low density of triplets i. e. for low enough
temperature. At high temperature, the triplet-triplet interactions
cannot be neglected. However, the mean-field theory gives nevertheless
a qualitatively correct picture of the development of Raman
intensity. 
The Raman intensity (see Appendix) is given by:
\begin{eqnarray}\label{eq:sa_raman}
{\rm Im} \chi _{R}(\omega )&=&C^2(\theta_I,\theta_S) \int dk \coth \left( \frac{\omega _{k}}{2k_{B}T} \right)
\left( \frac{\Delta _{k}}{\omega _{k}}\right) ^{2}(\delta (\omega -2\omega
_{k})\nonumber \\ &-&\delta (\omega +2\omega _{k})),  \label{imsusc}
\end{eqnarray}
\noindent where
\begin{equation}
C(\theta_I,\theta_S)=\frac 1 4\left[\gamma_\perp \sin \theta_I \sin
\theta_S -\frac{\gamma_\parallel J_\perp}{J_\parallel} \cos \theta_I
\cos \theta_S \right] \label{matrix_elem1d}.
\end{equation}

It measures, up to a matrix element, the
density of states of the triplet excitations. 
The Raman scattering spectra thus displays two peaks at its edges, 
the first one at energy $\omega
=2\omega_\pi=2\Delta _{s}$ corresponding to the bottom of the triplet
band, and the
second one at $\omega =2\omega_0$, corresponding to the top 
of the triplet band. Such peaks have been observed
experimentally\cite{sugai_raman_sr14cu24o41,sugai_raman_lacuo25,cavo_raman_spectra,srcuo_raman_spectra}.
Within mean field theory, 
it is possible to consider the effect of a non-zero temperature. For
$J_\perp/J=10$ the
temperature effect is negligible as long as $T<0.5J_\perp$. 
For larger $T$, there
is a decrease of $s$ that causes a decrease of the bandwidth of
triplet excitations (see Fig.~\ref{fig:temp_effect}). 
The effect of $C(\theta_I,\theta_S)$ is to give an explicit dependence
on the polarization of incoming and outcoming radiation to the Raman
intensity. However, it does not give rise to 
a shift of the peaks when the polarization is changed.  In the case
$\gamma_\perp/\gamma_\parallel=J_\perp/J_\parallel$, the Raman
intensity is proportional to $cos^2(\theta_I+\theta_S)$ as pointed out
in Ref.~\onlinecite{freitas_raman}. 

Eq. (\ref{imsusc}) was derived  neglecting triplet-triplet
scattering. Such scattering is responsible for the broadening of the
peaks in experiments\cite{sugai_raman_sr14cu24o41,sugai_raman_lacuo25,cavo_raman_spectra,srcuo_raman_spectra}. A full calculation of the effect of
triplet-triplet scattering is a very complex task. 
Physically, the most important effect of these interactions is 
that triplet bosons are not anymore exact eigenstates of the
Hamiltonian and acquire a finite lifetime $\tau$.
An estimate of this lifetime is given by the Fermi Golden rule as:

\begin{eqnarray}
\frac 1 \tau &=& 2\pi  \sum_{k',q} |V(q)|^2 n_B(k')
(1+n_B(k'+q))\nonumber \\
&\times &\delta(\epsilon(k+q)+\epsilon(k-q)-\epsilon(k)-\epsilon(k')).
\end{eqnarray}

The potential for triplet-triplet interactions being short
ranged, $V(q)\simeq V(q=0)=V_0$. Since the quasiparticle
dispersion is of the form: $\omega_k=\omega_{\pi} +
\frac{k^2}{2m_{\pi}}$ at the bottom of the spectrum,
 $\frac 1 \tau$ will be dominated by a factor
$e^{-\omega_\pi/T}$. The rest of the expression reduces to 
a phase space factor, independent of temperature.

The existence of a finite lifetime for the quasiparticles leads to the
replacement 
\begin{equation}
\delta(\omega-2\omega_k) \to \frac{\Gamma} \pi \frac 1 {(\omega-2\omega_k))^2
+ \Gamma^2}, 
\end{equation}
in Eq. \ref{eq:sa_raman}. At temperatures $T\ll \omega_0$, $\Gamma$ is
small and such replacement does not affect the intensity very much
except at the edges where it cuts the square root singularities of the
density of states. Thus, we can restrict to the consideration of Raman
intensity close to the edges of the spectrum. 

Let us first consider  the threshold $\omega\sim
\omega_\pi$. The Raman intensity is approximately:
\begin{eqnarray}
I(\omega)&=&\coth\left(\frac{\omega_\pi}{k_B T}\right)
\left(\frac{\Delta_\pi}{\omega_\pi}\right)^2 \int
\frac{dk}{2\pi}  \frac{\Gamma_\pi}{\pi} 
\frac 1 {(\omega-2\omega_\pi-\frac{k^2} {m_\pi})^2 +
\Gamma_\pi^2}\nonumber \\
& =& \coth\left(\frac{\omega_\pi}{k_B T}\right)
\left(\frac{\Delta_\pi}{\omega_\pi}\right)^2 \tilde{I}(\omega) 
\end{eqnarray}

where
\begin{equation} 
m_\pi=\frac{2 \sqrt{1-d}}{d \left(\frac {J_\perp}{4}-\mu\right)}
\end{equation}
The calculation leads to: 
 \begin{equation}\label{eq:raman_sa_damping}
\tilde{I}(\omega)=
\frac {\sqrt{\frac{m_\pi\Gamma_\pi^2}{8\pi^2}}}
{{\sqrt{(\omega-2\Delta)^2+\Gamma_\pi^2}}\sqrt{\sqrt{(\omega-2\Delta)^2+\Gamma_\pi^2}-\omega+2\Delta}}. 
\end{equation}

The resulting plot of $I(\omega)$ versus $\omega$ is shown on figure
\ref{fig:plot} and is qualitatively similar to the behavior of the
Raman intensity close to $2\Delta_s$ in experimental systems (see Fig. 3 in
Ref. \onlinecite{cavo_raman_spectra} or Fig. 5 in
Ref. \onlinecite{sugai_raman_sr14cu24o41}).  
Eq. (\ref{eq:raman_sa_damping}) can be cast in the scaling form:
  \begin{equation}
\tilde{I}(\omega)=\sqrt{\frac{8\pi^2\Gamma_\pi}{m_\pi}} I_{1D}(\omega)=
\frac 1 {\sqrt{x^2+1}\sqrt{\sqrt{x^2+1}-x}},
 \end{equation}
 
\noindent where
$x=\frac{\omega-2\Delta}{\Gamma_\pi}$. The corresponding plot is
Fig.~\ref{fig:univ1d}. It would be interesting to determine whether
the experimental data satisfy such scaling form.  
 A similar calculation can be
done for $\omega \sim \omega_0$, with result: 
 \begin{eqnarray}\label{eq:raman_sa_upper}
I(\omega)=\coth\left(\frac{\omega_0}{2k_B
T}\right)\left(\frac{\Delta_0}{\omega_0}\right)^2 \sqrt{\frac{m_0}{8
\pi^2 \Gamma_0}}I_{1D}\left(\frac{2\omega_0-\omega} {\Gamma_0}\right)
\end{eqnarray}
where:
\begin{equation} 
m_0=\frac{2 \sqrt{1+d}}{d \left(\frac {J_\perp}{4}-\mu\right)}
\end{equation}

Although in experiments on $\mathrm{Sr_{14}Cu_{24}O_{41}}$ the
condition $J_\perp \gg J$ is certainly not 
satisfied, it is nevertheless interesting to compare the dependence
predicted by Eq. (\ref{eq:raman_sa_upper}) with the two magnon
intensity of Fig. 4 in \onlinecite{sugai_raman_sr14cu24o41} or Fig. 7
of Ref. \onlinecite{popovic_raman_sr14cu24o41}.  
At temperatures small in comparison to the spin gap, the damping
effect is the dominant feature near the edges. At low temperature
there is a  rapid variation of
the damping rate $\Gamma=1/\tau\sim e^{-\Delta_s/T}$ with temperature 
and thus a rapid variation of the intensity. It would be interesting
to check whether such dependence of the damping rate is consistent
with experiments. As a final note, we would like to point out that
Equations (\ref{eq:raman_sa_damping}) and (\ref{eq:raman_sa_upper}) do
not lead in general to a symmetric Raman spectra. In particular, in
the case of $\mathrm{CaV_2O_5}$, a different damping rate and 
different masses of excitations could
explain the difference of behavior of the intensity at the two
edges\cite{cavo_raman_spectra}.  

\section{Coupled ladders}\label{sec:coupled_ladders}
\subsection{Mean field equations}
In this section we start to consider the effects on the Raman
intensity due to interladder interaction. Our final aim is to derive
an expression of  the Raman intensity in the case of the array of
ladders whose Hamiltonian was given in Eq. (\ref{hladder_array}). Such
ladder array model is of interest for systems such as
$\mathrm{KCuCl_3}$ \cite{cavadini_kcucl3_neutrons} or possibly
CuHpCl\cite{hammar_transition_cuhpcl} according to some neutron
scattering experiments. 
As a simpler problem,
 we will also discuss a double rung
ladder in which two spin-1/2 ladders are connected by interladder
exchange of strength $\lambda
^{\prime }J_{\perp }$, that is weaker than the intraladder coupling. 
In the BOT the Hamiltonian
given by Eq. \ref{hamda} where the index $j$ can run now from 1 to N.
 First we apply the
Green's function method to determine the spectrum, 
whose complete calculation is
reported in the Appendix  . It will be given by

\begin{eqnarray}
\omega^2(k_x,k_y)&=&\left(\frac{J_\perp} 4 -\mu -J_\parallel s^2
\cos(k_x ) + \frac{\lambda' J_\perp s^2 } 2 \cos (k_y ) \right)^2
\nonumber \\ 
&-& \left(J_\parallel s^2 \cos(k_x ) - \frac{\lambda' J_\perp s^2 } 2 \cos (k_y )\right)^2 .
\end{eqnarray}

The self-consistent equations in the saddle-point
approximation that permit us to determine the parameters ($s^{2},\mu $) at
T=0 are given by
\begin{eqnarray}
(s^{2}-\frac{3}{2})+\int_{-\pi /a}^{\pi /a}\frac{dk_{x}}{(2\pi )}\int_{-\pi
/a}^{\pi /a}\frac{dk_{y}}{(2\pi )}\left[ \frac{1-\frac{d}{2}f(k_{x},k_{y})}{%
\sqrt{1-df(k_{x},k_{y})}}\right] &=&0  \label{selfcon1_array} \\
(\frac{3}{2}+2\frac{\mu }{J_{\perp }})+\lambda \int_{-\pi /a}^{\pi /a}\frac{%
dk_{x}}{(2\pi )}\int_{-\pi /a}^{\pi /a}\frac{dk_{y}}{(2\pi )}\left[ \frac{%
f(k_{x},k_{y})}{\sqrt{1-df(k_{x},k_{y})}}\right] &=&0  \label{selfcon2_array}
\end{eqnarray}
where

\begin{eqnarray}
d &=&\frac{\lambda s^{2}}{(\frac{1}{4}-\frac{\mu }{J_{\perp }})} \\
f &=&(\frac{2\lambda}{\lambda'}\cos k_{x}-\cos k_{y})
\label{param_array}
\end{eqnarray}

The equations (\ref{selfcon1_array}) and (\ref{selfcon2_array}) can be combined to
obtain the following single mean-field equation that is given by

\[
\frac{d}{\lambda}=\frac{3}{2}-\int \int \frac{d^{2}k}{(2\pi )^{2}}\text{ }%
\frac{1}{\sqrt{1-df}}.
\]

Once we have determined $d$, $s^2$ and the chemical potential $\mu$,
we are ready to perform the analysis of Raman susceptibility.

In the case of the double ladder, the energy dispersion is determined
by restricting $k_y$ to take the discrete values $0$ or $\pi$. This
leads to two branches in the spectrum due to the fact
that as soon as $\lambda ^{\prime }$ is different from zero the
triplet excitations not
only delocalize along the single ladders but also across the ladders. The
two branches are usually called the bonding (the one below) and antibonding
(the one above) band.

The resulting excitation spectrum is given by\cite{gopalan_2ch}:
\begin{eqnarray}
\omega^2_{\pm}(k)&=&\left( \frac{J_{\perp }}{4}-\mu +J_{\parallel
}s^{2}\cos (k)\pm \frac{J_{\perp }s^{2}}{2}\right) ^{2} \nonumber \\ 
&-&\left( J_{\parallel
}s^{2}\cos (k)\mp \frac{J_{\perp }s^{2}}{2}\right) ^{2}
\end{eqnarray}

The chemical potential $\mu$ and the parameter $s$ must obtained as above
by solving the saddle-point equations, with the integral over $k_y$
changed to a discrete summation.  They read:

\begin{eqnarray}
(s^{2}-\frac{3}{2})+\frac{1}{4}\int_{-\pi /a}^{\pi /a}\frac{dk}{(2\pi )}%
\left[ \frac{1+\frac{d}{2}f_{+}}{\sqrt{1+df_{+}}}+
\frac{1+\frac{d}{2}f_{-}}{%
\sqrt{1+df_{-}}}\right]  &=&0 \nonumber \\  \label{selfcon1} \\
(\frac{3}{2}+2\frac{\mu }{J_{\perp }})-\frac{\lambda }{2}\int_{-\pi /a}^{\pi
/a}\frac{dk}{(2\pi )}\left[ \frac{f_{+}}{\sqrt{1+df_{+}}}+
\frac{f_{-}}{\sqrt{%
1+df_{-}}}\right]  &=&0  \nonumber \\ \label{selfcon2}
\end{eqnarray}
where

\begin{eqnarray}
& & d =\frac{\lambda s^{2}}{(\frac{1}{4}-\frac{\mu }{J_{\perp }})} \\
& & f_{\pm} =(\cos (k)\pm \frac{1}{2}\frac{\lambda ^{\prime }}{\lambda })
\label{param}
\end{eqnarray}

\noindent Eqs.  (\ref{selfcon1}) and (\ref{selfcon2}) can be combined\cite{gopalan_2ch} to
obtain the following single mean-field equation

\begin{eqnarray}
\frac{d}{\lambda }&=&3-\frac{2}{\pi }\frac{1}{\sqrt{1+d(1+a)}}K\left( \frac{2d%
}{1+d(1+a)}\right) \nonumber \\
 &-&\frac{2}{\pi }\frac{1}{\sqrt{1+d(1-a)}}K\left( \frac{2d}{%
1+d(1-a)}\right) ,
\end{eqnarray}
where $a=$ $\frac{1}{2}\frac{\lambda ^{\prime }}{\lambda }.$

At this point it is worth analyzing the spin-triplet gap $\Delta _{s}$. It is given by

\[
\Delta _{s}=\left( \frac{J_{\perp }}{4}-\mu \right) \left[ 1-d\left( 1+\frac{%
1}{2}\frac{\lambda ^{\prime }}{\lambda }\right) \right] ^{1/2}.
\]
It was shown by Gopalan et al. \cite{gopalan_2ch} that as $\lambda'$
was increased, the spin gap was reducing. 

We can now turn to the calculation of the Raman susceptibility.

\subsection{Derivation of the Raman Susceptibility}

For the calculation of the Raman susceptibility for coupled ladders 
we make use of
the expression (\ref{generalr}) for the Raman operator.
To start with, we find that:
\begin{eqnarray}
\langle T_\tau H_R(\tau) H_R(0)\rangle \nonumber \\ =  3 \sum_{{\bf k}}\left\{ A^2({\bf k})
\left[ G_{\bf k} (\tau)G_{\bf k} (-\tau)+ F_{\bf k} (\tau)
(F^\dagger)_{\bf k}(-\tau) \right]\right. \nonumber \\
+ 2A({\bf k}) B({\bf k}) \cos (k_x
a) \left[G_{\bf k} (\tau)(F^\dagger)_{\bf k}(-\tau)+G_{\bf k}
(-\tau)(F^\dagger)_{\bf k}(\tau)\right . \nonumber \\
 +  \left . G_{\bf k} (-\tau) F_{\bf k} (\tau) +G_{\bf k} (\tau) F_{\bf k} (-\tau)
\right]\nonumber \\
+  B^2({\bf k}) \left[e^{2ik_x a} (F^\dagger)_{\bf
k}(-\tau)(F^\dagger)_{\bf k}(-\tau) + G_{\bf k} (\tau) G_{\bf k}
(\tau) \right. \nonumber \\
+  \left. \left. e^{-2ik_x a} F_{\bf k} (\tau) F_{\bf k} (\tau) +G_{\bf
k} (-\tau)G_{\bf k} (-\tau)\right] \right\}.
\end{eqnarray}
Where $G,F$ represent the normal and anomalous Green's Functions of
the triplet bosons and are defined in the Appendix. 
The term in $A^2({\bf k})$ has already been calculated in the single ladder
problem, while the remaining terms give new contributions to the Raman
intensity. Going to Fourier space and making use of the Matsubara summations reported
in the Appendix, we find that the Raman response function is:

\begin{equation}
\chi_R(\i \omega_n)= E_I^2 E_S^2 \sum_{\bf k} M_{\bf k} \coth \left(\frac{\beta \omega_{\bf
k}} 2\right) \left[ \frac 1 {i\omega_n -2 \omega_k} -\frac 1 {i\omega_n +
2\omega_k} \right],
\label{chigeneral}
\end{equation}

\noindent where:

\begin{eqnarray}
\label{mk}
M_{\bf k}=4B_{\bf k}^2\cos^2 (k_xa) \left(\frac{\Delta_{\bf k}}{\omega_{\bf k}} \right)^2
+ 2 A_{\bf k}B_{\bf k} \cos (k_x a) \frac{\Lambda_{\bf k} \Delta_{\bf
k}}{\omega_{\bf k}^2}\nonumber \\ -2 A_{\bf k}^2  \left(\frac{\Delta_{\bf
k}}{\omega_{\bf k}} \right)^2. 
\end{eqnarray}

In particular, if $\gamma_\parallel /J_\parallel =\gamma_\perp /J_\perp =\gamma'_\perp /J'_\perp$, due to
relation (\ref{akvsbk}), the expression for $M_{\bf k}$ simplifies:

\begin{equation}
\label{akbk}
M_{\bf k}=A_{\bf k}^2
\left ( \frac{\Lambda_{\bf k} \Delta_{\bf k}}{\omega_{\bf k}^2} -
\left(\frac{\Delta_{\bf k}}{\omega_{\bf k}} \right)^2\right ),
\end{equation}

\noindent where we have used the following notation:
\begin{eqnarray}
&& \Lambda_{\mathbf k}=(\frac{J_\perp} 4 -\mu) -\left (J_\parallel s^2 \cos (k_x) -
\frac{\lambda' J_\perp s^2 } 2 \cos (k_y )\right ), \nonumber \\
&& (2\Delta_{\mathbf k})= J_{\parallel} s^2 \cos (k_x) -\frac{\lambda' J_\perp s^2}{2}\cos (k_y ).
\end{eqnarray}
 Note that the two
coupled ladders case is included in the formula above since we can
restrict the sum over $k_y$ to $0$ and $\pi$.
The expression (\ref{chigeneral}) is very general and
contains an explicit dependence on the polarization and the electric field
intensity. We emphasize that the peak position predicted by
Eq.~(\ref{chigeneral}) only depends on the spectrum
of the ladder system. In particular, it is completely independent 
of the polarization
of ingoing as well as outgoing radiation.

Performing an analytic continuation,  the intensity of the Raman spectrum is given by

\begin{equation}
{\rm Im}\chi _{R}(\omega )=\int \int  \frac{d^2k}{(2\pi)^2}M_{\bf k}(\delta (\omega -2\omega _{k})-\delta
(\omega +2\omega _{k,})),  \label{imsusc_array}
\end{equation}

\noindent where $M_{\bf k}$ is given by (\ref{akbk}).
Since the $\gamma$'s appearing in the expression for $M_{\bf k}$
can be treated as phenomenological parameters we will perform the
explicit calculation for
$\gamma_\parallel /J_\parallel =\gamma_\perp /J_\perp =\gamma'_\perp /J'_\perp$.

\subsubsection{The Raman intensity of the two-rung ladder}

Using the previous formulas, the Raman intensity for the two-rung ladder will be
obtained  from formula (\ref{imsusc_array}) reducing the integral on $k_y$ to a sum
for  $k_y=0$ and $k_y=\pi$:

\begin{eqnarray}
{\rm Im}\chi _{R}(\omega )&=&\sum_{\alpha =1,2}\int dk \left( \frac{A_k^2
\left (\Lambda_k\Delta_k-\Delta _{k}^2\right )}{\omega
_{k,\alpha }^2}\right)(\delta (\omega -2\omega _{k,\alpha })\nonumber
\\
&-&\delta (\omega +2\omega _{k,\alpha })).  \label{imsusc_2lad}
\end{eqnarray}

\noindent Here we have summed on the two triplet bands. Performing the integral
in (\ref{imsusc_2lad}), we obtain as a final result

\begin{eqnarray}
{\rm Im}\chi _{R}(\omega )=\nonumber \\ A_0\sum_{\alpha=\pm} \frac{
\left| \left( \frac{\omega
}{2\left( J_{\perp }/4-\mu \right) }\right) ^{2}-a_{\alpha }\right|^3
\left[1+\frac{1}{4}\left[ \left( \frac{\omega }
{2\left( J_{\perp }/4-\mu \right) }\right) ^{2}-a_{\alpha } \right] \right]}{4\omega
\sqrt{\left( \frac{2Js^{2}}{J_{\perp }/4-\mu }\right) ^{2}-\left[ \left( \frac{\omega
}{2\left( J_{\perp }/4-\mu \right) }\right) ^{2}-a_{\alpha }\right] ^{2}}},
\end{eqnarray}

\noindent where  $A_0=\left( J_{\perp }/4-\mu \right)^3
\left (\frac{\gamma_\parallel}{J_\parallel}\right )^2\cos^2(\theta_I+\theta_S)$ , $a_{\pm }=(1\pm \frac{1}{2}\frac{\lambda ^{\prime }}{\lambda }).$
As shown in Fig.\ref{fig:intensityda}, the Raman spectrum shows four
peaks in correspondence of the bottom and
the top the bonding and antibonding bands. From simple density of states
argument, no signal should be seen for $\omega <2\Delta _{s},$ where $%
\Delta _{s}$ is the singlet-triplet gap.

\subsubsection{The array of ladders}
\label{sec:two_dimensional}
In this case we start from equation (\ref{imsusc_array}), 
by introducing the two-dimensional density of states $\rho(\omega)$ 
 and  we can write
\begin{equation}
{\rm Im}\chi _{R}(\omega )\propto \frac{1}{4}\rho(\omega/2)\lbrack \left (
\frac{\omega}{2(\frac{J_{\perp}}{4}-\mu)}\right )^2-1 \rbrack, \label{imsuscr_array}
\end{equation}

\begin{eqnarray}
\rho(\omega)=\frac{2}{\pi}\lbrack \frac{\omega}{(\frac{J_{\perp}}{4}-\mu)} \rbrack
\frac{1}{J_{\parallel}s^2}\nonumber \\
\int \frac{dk_y}{2\pi} \frac{1} {\sqrt{1-\left(
\frac{1}{d}\left( 1-\left(\frac{\omega}{(\frac{J_{\perp}}{4}-\mu)}\right)^2 \right)
+\frac{\lambda'}{2\lambda}\cos k_y \right)^2 } }.
\end{eqnarray}

The last integral is performed for
$Max(-1,+\frac{2\lambda}{\lambda'}-\frac{2\lambda}{d\lambda'}\left ( 1-\left
(\frac{\omega}{(\frac{J_{\perp}}{4}-\mu)} \right )^2 \right ))<\cos
k_y<Min(1,-\frac{2\lambda}{\lambda'}-\frac{2\lambda}{d\lambda'}\left ( 1-\left
(\frac{\omega}{(\frac{J_{\perp}}{4}-\mu)} \right )^2 \right ))$. The results show that
the intensity is non-zero only between the maximum and the minimum of $\omega
({\mathbf{k}})$ and a discontinuity appears at the band edge. The
behavior of the 2D Raman intensity is sketched on  Fig.~\ref{fig:intensity2d}.

 In two dimensions, we can expect
two types of singularities in $\rho(\omega)$, namely
discontinuities and logarithmic
singularities\cite{ziman_solid_book}. Close to these singularities,
damping effects due to triplet-triplet scattering will play an
important role, in analogy to the single ladder problem.  
In the present case
the two-magnon continuum starts at energy:
$\omega_{-}=2\Delta_s=2(J_{\perp}/4-\mu)\sqrt{1-d(1+\frac{\lambda}{2\lambda'})}$,
the other edge of the Raman spectrum being at
$\omega_{+}=2(J_{\perp}/4-\mu)\sqrt{1+d(1+\frac{\lambda}{2\lambda'})}$.
No magnetic Raman scattering is observed for $\omega \not \in
[\omega_{-},\omega_{+}]$.  At these points, discontinuities in the
Raman intensity would appear in the absence of damping.
Taking damping into account close to the edges of the spectrum, we find
that the intensity is of the form: 

\begin{eqnarray}
I_{2D}^{\text{edge}}(\omega)=\int \frac{k dk}{2\pi}
\frac{\Gamma}{(\omega-2\Delta-k^2)^2
+\Gamma^2}\nonumber \\ =\frac{2}{\pi^2}\left(\frac \pi 2 + \arctan
\frac{\omega-2\Delta}{\Gamma}\right).
\end{eqnarray}
The steps at the edges of the spectrum are thus smoothed out by
damping of excitations. The corresponding plot is Fig.~\ref{fig:univ2d}.

At a logarithmic singularity $\omega\sim 2(J_{\perp}/4-\mu)\sqrt{1\pm
d(1-\frac{\lambda}{2\lambda'})}$, we find that:
\begin{equation}
I_{2D}^{\text{peak}}(\omega)=\int \frac{dk_x dk_y}{(2\pi)^2} \frac{\Gamma}{\pi} \frac 1
{(\omega-2\Delta-k_x^2/m+k_y^2/m)^2 +\Gamma^2}, 
\end{equation}

\noindent and thus:
\begin{equation}
I(\omega)\sim \ln \frac{\Lambda^2}{\Gamma^2+(\omega-2\Delta)^2},
\end{equation}
\noindent where $\Lambda$ is an energy cutoff. Again, the height of the peaks is
limited by the damping of excitations. Note that in contrast to the
one dimensional case, the peaks are not obtained at the edges of the
two magnon Raman spectrum but are superposed onto a positive
two-magnon background. Only in the one dimensional 
limit do the peaks merge with the steps at
the edge of the spectrum. Also, in the two dimensional case, the peaks
are symmetric (see Fig.~\ref{fig:logpeaks}) in contrast to the 1D
case. 
Thus, Raman scattering can prove useful to
discriminate quasi-one dimensional systems from quasi two-dimensional
systems with a spin
gap~\cite{hammar_transition_cuhpcl,chaboussant_nmr_diaza} and provide
bounds on interladder coupling.   

\section{Conclusion}

We have presented an analysis of the Raman spectra for various coupled spin
 ladder systems based on the Loudon-Fleury photon-induced
super-exchange theory, representing the spins using  the Bond Operator Formalism. For a single ladder, we have shown that peaks should appear 
at the edges of the Raman spectrum as has been observed in
experiments\cite{cavo_raman_spectra,popovic_raman_sr14cu24o41}. 
The shape of these peaks is
determined by the damping of triplet excitations due to interactions. 
A scaling plot of the Raman intensity versus frequency should permit
to determine the lifetime of excitations in the ladder. Simple
arguments lead us to expect that the lifetime of excitations must vary
as $e^{\Delta/T}$ where $\Delta$ is the spin gap. Another consequence
of a positive temperature on the spin ladder is the narrowing of the
spectrum as temperature is increased. It would be interesting to
compare our predictions to the available experimental data on spin
ladder systems. In the case of the coupled array of ladders, we have
shown that the Raman scattering intensity measures the density of
states of triplet excitations. Thus, the singularities of the Raman
intensity should indicate very clearly whether the system is one or
two dimensional. In both case, the peak position was independent from
the polarization of the incoming and scattered radiation. 
It would be interesting to generalize our
analysis to other two dimensional spin gap systems such as the models
with orthogonal dimers\cite{ueda_ortho_dimers} that can be used to model
$\mathrm{SrCu_2(BO_3)_2}$. Another direction in which our work could
be extended is to conducting spin
ladders\cite{sugai_raman_sr14cu24o41} in which case charge degrees of
freedom could also come into play. 
 
\acknowledgments 
We thank G. Blumberg, R. Chitra, T. Giamarchi and C. M. Varma  for
illuminating discussions. E. O. acknowledges partial 
support from NSF under grants DMR 96-14999 and DMR 99-76665. R. Citro
acknowledges support from 
INFM and thanks Laboratoire de Physique Th\'eorique de l'\'Ecole
Normale Sup\'erieure for kind hospitality. 

%\newpage

\appendix

\section*{Green's Functions method for the array of ladders}\label{app:greens}

We introduce the following  Green's functions:
\begin{eqnarray}
G_{\alpha}({\bf k},\tau) =-\langle T_\tau t_{{\bf k},\alpha} (\tau)
t^\dagger_{{\bf k},\alpha}(0) \rangle  \nonumber \\
\tilde{G}_{\alpha}({\bf k},\tau) =-\langle T_\tau t^\dagger_{{\bf k},\alpha} (\tau)
t_{{\bf k},\alpha}(0) \rangle  \nonumber \\
F_{\alpha}({\bf k},\tau) =-\langle T_\tau t_{{\bf k},\alpha} (\tau) t_{-{\bf k},\alpha}(0)
\rangle  \nonumber \\
(F^\dagger)_{\alpha}({\bf k},\tau)=-\langle T_\tau t^\dagger_{{\bf k},\alpha}(\tau)
t^\dagger_{{\bf k},\alpha}(0) \rangle
\end{eqnarray}

The equations of motion permit us to obtain the following coupled equations:
\begin{eqnarray}
-i \omega_n G_\alpha({\bf k},\omega_n) = -1 -\left(\frac{J_\perp} 4
-\mu\right) G_\alpha({\bf k},\omega_n) \nonumber \\
- s^2 \left( J_\parallel \cos(k_x )
- \frac{\lambda' J_\perp} 2 \cos (k_y ) \right) \left[G_\alpha({\bf k},%
\omega_n) + F^\dagger_\alpha({\bf k},\omega_n)\right]  \nonumber \\
-i \omega_n F^\dagger_\alpha({\bf k},\omega_n) =  \left(\frac{J_\perp} 4
-\mu\right)F^\dagger_\alpha({\bf k},\omega_n)\nonumber \\
 + s^2 \left(J_\parallel
\cos(k_x ) - \frac{\lambda' J_\perp} 2 \cos (k_y ) \right) \left[%
G_\alpha({\bf k},\omega_n) + F^\dagger_\alpha({\bf k},\omega_n)\right]
\nonumber \\
\end{eqnarray}
and:
\begin{eqnarray}
-i \omega_n \tilde{G}_\alpha({\bf k},\omega_n) = 1 +\left(\frac{J_\perp} 4
-\mu\right) G_\alpha({\bf k},\omega_n) \nonumber \\
 + s^2 \left( J_\parallel \cos(k_x )
- \frac{\lambda' J_\perp} 2 \cos (k_y ) \right) \left[\tilde{G}_\alpha({\bf k},%
\omega_n) + F_\alpha({\bf k},\omega_n)\right]  \nonumber \\
-i \omega_n F_\alpha({\bf k},\omega_n) =  -\left(\frac{J_\perp} 4
-\mu\right)F^\dagger_\alpha({\bf k},\omega_n)\nonumber \\
 - s^2 \left(J_\parallel
\cos(k_x ) - \frac{\lambda' J_\perp} 2 \cos (k_y ) \right) \left[%
\tilde{G}_\alpha({\bf k},\omega_n) + F_\alpha({\bf k},\omega_n)\right]
\nonumber \\
\end{eqnarray}

It is convenient to introduce:
\begin{equation}
\Lambda({\bf k})=\frac{J_\perp} 4 -\mu +J_\parallel s^2 \cos(k_x ) -
\frac{ J'_\perp s^2 } 2 \cos (k_y )
\end{equation}
and:
\begin{equation}
2\Delta({\bf k})=J_\parallel s^2 \cos(k_x
) - \frac{ J'_\perp s^2 } 2 \cos (k_y)
\end{equation}

so that:
\begin{eqnarray}
G({\bf k},\omega_n)=(\tilde{G}({\bf k},\omega_n))^*=-\frac{i\omega_n +\Lambda({\bf k})}{\omega_n^2 +
\omega({ \bf k})^2} \nonumber \\
F({\bf k},\omega_n)=F^\dagger({\bf k},\omega_n)=\frac{2\Delta_k}{\omega_n^2 +
\omega({ \bf k})^2} 
\end{eqnarray}
\noindent Where we have:
\begin{eqnarray}
\omega^2({\bf k})=\Lambda^2({\bf k})-(2\Delta({\bf k}))^2
\end{eqnarray}
The dispersion for the isolated ladder is obtained by taking
$J'_\perp=0$ in the preceding formulas.

A convenient decomposition of the Green's functions is:
\begin{eqnarray}
G({\bf k},\omega_n)=\frac 1 2 \left[\left(1 + \frac{\Lambda({\bf
k})}{\omega({\bf k})}\right) \frac{1}{i\omega_n +\omega({\bf
k})}\right. \nonumber \\ \left.
+\left(1 - \frac{\Lambda({\bf k})}{\omega({\bf k})}\right)
\frac{1}{i\omega_n -\omega({\bf k})}\right] \nonumber \\
F^\dagger(k,\omega_n)=\frac{\Delta({\bf k})}{\omega({\bf
k})}\left[\frac{1}{i\omega_n +\omega({\bf k})}-\frac{1}{i\omega_n
-\omega({\bf k})}\right]
\end{eqnarray}

Using this decomposition, the Matsubara sums are reduced to:
\begin{eqnarray}
\frac 1 \beta \sum_{\omega_n} \frac{1}{i(\omega_n + \nu_n)\pm\omega({\bf
k})} \frac 1 {i \omega_n \pm \omega({\bf
k})} =0 \\
\frac 1 \beta \sum_{\omega_n} \frac{1}{i(\omega_n - \nu_n)\pm \omega({\bf
k})} \frac 1 {i \omega_n \mp \omega({\bf
k})} =\pm \frac{\coth\left(\frac{\beta \omega_{\bf k}}{2}\right)}{i
\nu_n \mp 2 \omega_{\bf k}}\\
\end{eqnarray}
with the appropriate factor.

\noindent This gives us the following integrals:
\begin{eqnarray}\label{eq:m_sums}
 \frac 1 \beta \sum_{i \nu_n} G_{\bf k}(i\nu_n) G_{\bf k}(i\nu_n -i
\omega_n)=\nonumber \\
 -\left(\frac{\Delta_{\bf k}}{\omega_{\bf k}}\right)^2 \coth
\left(\frac{\beta \omega_{\bf k}}{2}\right) \left[ \frac 1 {i \omega_n
-2 \omega_{\bf k}} -  \frac 1 {i \omega_n
+2 \omega_{\bf k}} \right]\nonumber \\
 \frac 1 \beta \sum_{i \nu_n} F_{\bf k}(i\nu_n) F_{\bf k}(i\nu_n -i
\omega_n)= \nonumber \\ -\left(\frac{\Delta_{\bf k}}{\omega_{\bf k}}\right)^2 \coth
\left(\frac{\beta \omega_{\bf k}}{2}\right) \left[ \frac 1 {i \omega_n
-2 \omega_{\bf k}} -  \frac 1 {i \omega_n
+2 \omega_{\bf k}} \right]\nonumber \\
\frac 1 \beta \sum_{i \nu_n} [F_{\bf k}(i\nu_n) G_{\bf k}(i\nu_n -i
\omega_n) + G_{\bf k}(i\nu_n) F_{\bf k}(i\nu_n -i
\omega_n)] \nonumber \\ 
= \frac{\Delta_{\bf k}\Lambda_{\bf k}}{\omega_{\bf k}^2} \coth
\left(\frac{\beta \omega_{\bf k}}{2}\right) \left[ \frac 1 {i \omega_n
-2 \omega_{\bf k}} -  \frac 1 {i \omega_n
+2 \omega_{\bf k}}\right] \nonumber \\
\frac 1 \beta \sum_{i \nu_n} G_{\bf k}(i\nu_n) [G_{\bf k}(i\nu_n -i
\omega_n)+ G_{\bf k}(-i\nu_n +i
\omega_n)]\nonumber \\
=\left(\frac{\Delta_{\bf k}}{\omega_{\bf k}}\right)^2
\coth \left(\frac{\beta \omega_{\bf k}}{2}\right)  \left[ \frac 1 {i \omega_n
+2 \omega_{\bf k}}
- \frac 1 {i \omega_n
-2 \omega_{\bf k}} \right]
\end{eqnarray}

This allows the complete calculation of the Raman intensity. For
instance, in the single rung case, the intensity is given by:

\begin{eqnarray}
\chi _{R}(i \omega _{n})&=&\frac 1 \beta C^2(\theta_I,\theta_S) \sum_{\nu _{n}}\int \frac{dk}{2\pi }%
\left[ G(k,i\nu _{n})G(k,i\nu _{n}-i\omega
_{n})\right. \nonumber \\  &+ &\left. F(k,i\nu _{n})F^{\dagger
}(k,i\omega _{n}-i\nu _{n})\right] ,  \label{eq:raman_gf}
\end{eqnarray}
\noindent where $C(\theta_I,\theta_S)$ is given by
Eq.~(\ref{matrix_elem1d}). 

Using Eqs. (\ref{eq:m_sums}) one obtains the expression
Eq. (\ref{eq:sa_raman}) for the Raman intensity.

\bibliographystyle{prsty}
%\bibliography{totphys,raman,spinchiral2,suthreelast}

\begin{figure}
\centerline{\epsfig{file=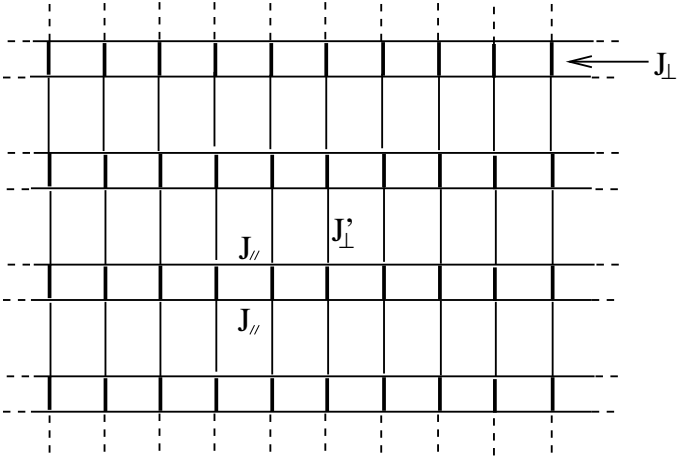,angle=0,width=9cm}} 
\caption{The two dimensional array of ladders.} \label{fig:array}
\end{figure}

\begin{figure}
\centerline{\epsfig{file=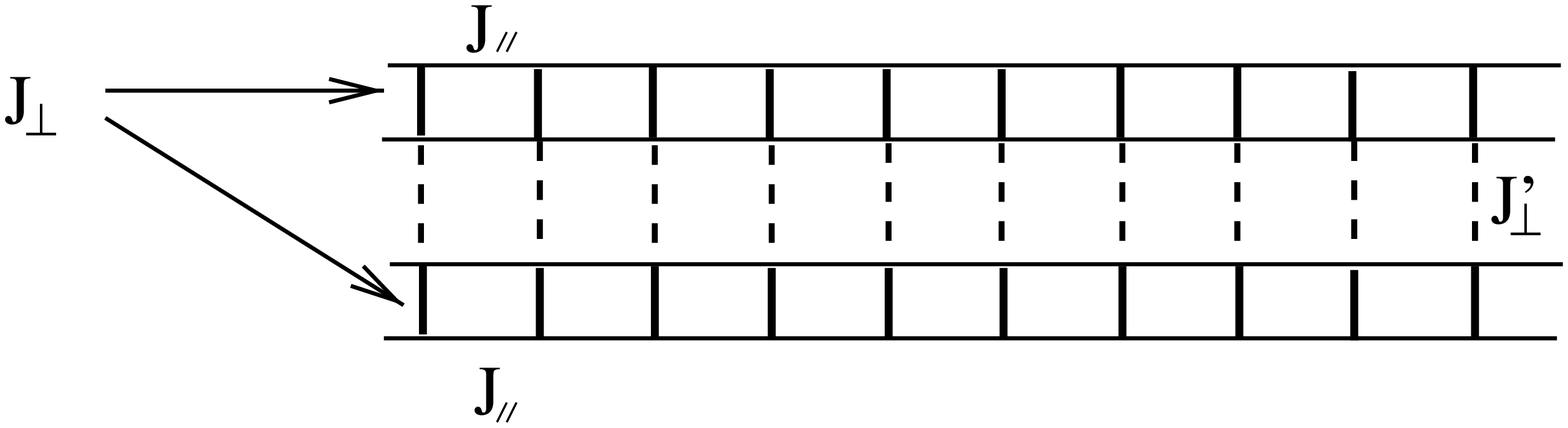,angle=0,width=9cm}}
\caption{The double ladder}
\label{fig:double_ladder}
\end{figure}

\begin{figure}
\centerline{\epsfig{file=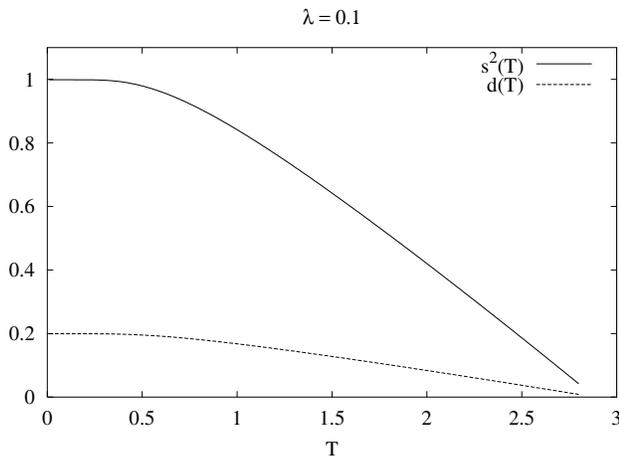,angle=-90,width=9cm}} 
\caption{The singlet order
parameter $s^2(T)$ and $d^2(T)$ as a function of the temperature (in units of $J_{\perp}$) for
$\lambda=0.1$.} \label{fig:sdvst}
\end{figure}

\begin{figure}
\centerline{\epsfig{file=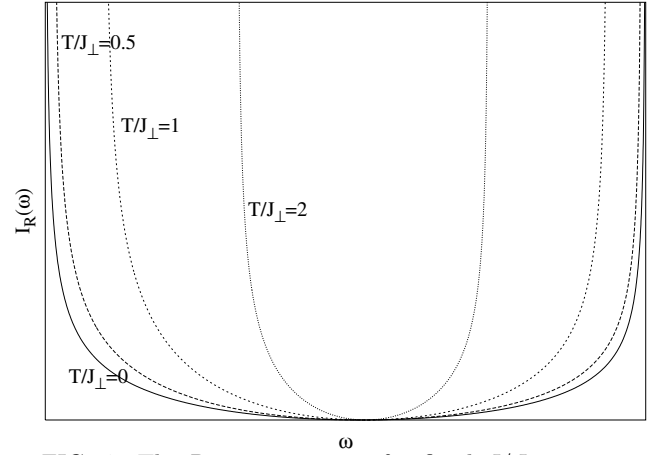,angle=-90,width=9cm}}
\caption{The Raman intensity for fixed $J/J_\perp=0.1$ and increasing
$T/J_\perp$. For $T/J_\perp<0.1$ the effect of temperature on the
Raman intensity is not visible. For $T/J_\perp>0.5$, the peaks are
moving toward each other as a consequence of the diminution of $s$ and
$\Delta$}
\label{fig:temp_effect}
\end{figure}

\begin{figure}
\centerline{\epsfig{file=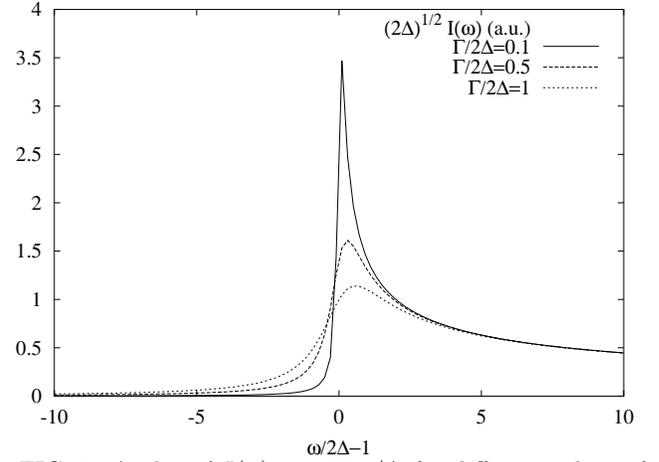,angle=-90,width=9cm}}
\caption{A plot of $I(\omega)$ versus $\omega/\Delta$
for different values of $\Gamma/\Delta$. Increasing temperature amounts
to increasing $\Gamma$ and results in a reduction of peak intensity
and the apparition of a small intensity below the gap}
\label{fig:plot}
\end{figure}

\begin{figure}
\centerline{\epsfig{file=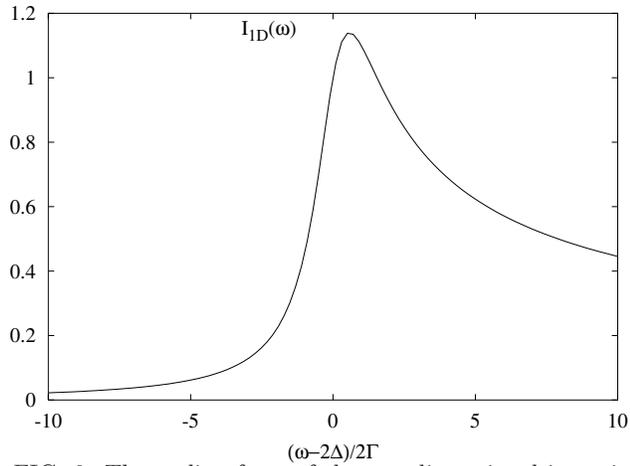,angle=-90,width=9cm}}
\caption{The scaling form of the one dimensional intensity
Eq.~(\ref{eq:raman_sa_damping})} 
\label{fig:univ1d}
\end{figure}

\begin{figure}
\centerline{\epsfig{file=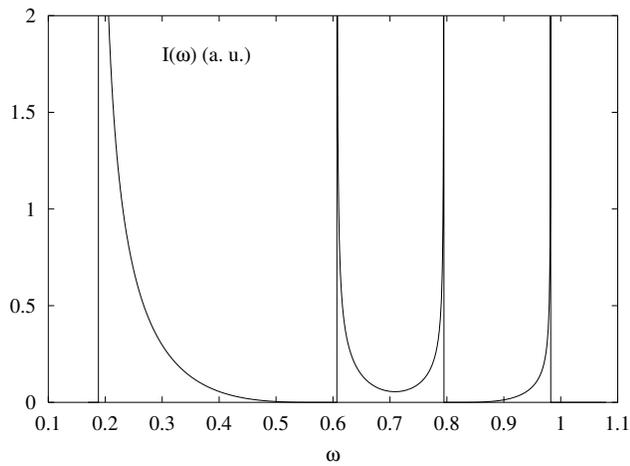,angle=-90,width=9cm}} 
\caption{Raman intensity for the two coupled ladders, for
$J/J_\perp=0.3,J'/J_\perp=0.1$.} 
\label{fig:intensityda}
\end{figure}

\begin{figure}
\centerline{\epsfig{file=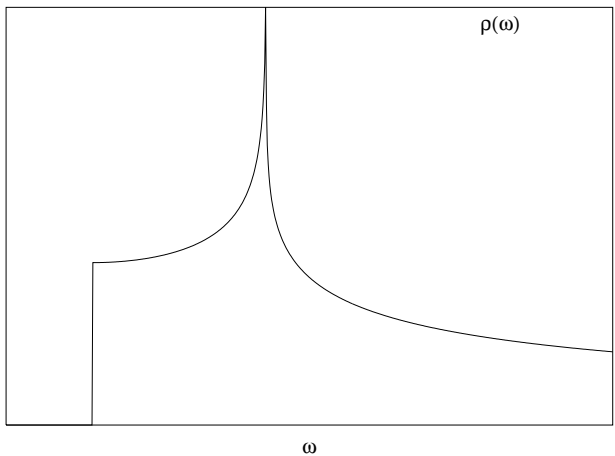,angle=-90,width=9cm}}
\caption{The behavior of the density of states in a two dimensional
system, showing a discontinuity at threshold and a logarithmic peak inside
the spectrum.} \label{fig:intensity2d}
\end{figure}

\begin{figure}
\centerline{\epsfig{file=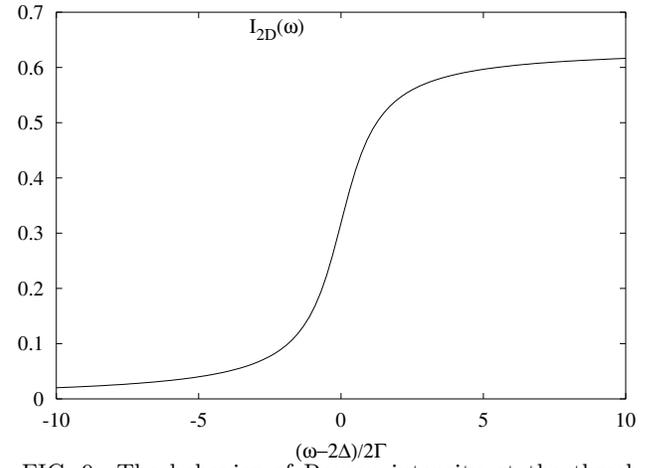,angle=-90,width=9cm}}
\caption{The behavior of Raman intensity at the threshold of the
two-magnon spectrum in a two dimensional ladder array taking into
account the damping of triplet excitations.}
\label{fig:univ2d}
\end{figure}

\begin{figure}
\centerline{\epsfig{file=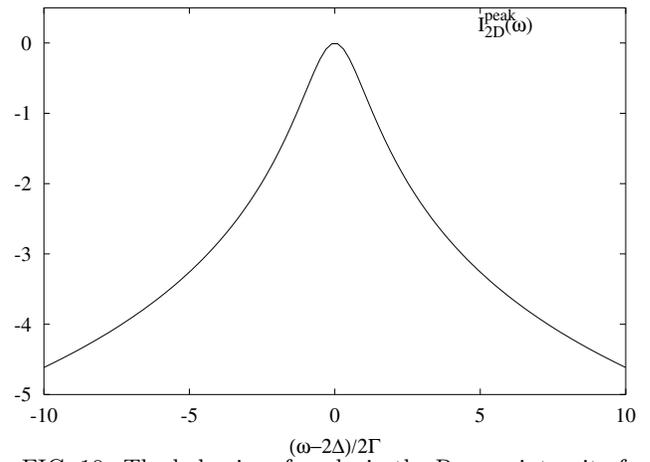,angle=-90,width=9cm}}
\caption{The behavior of peaks in the Raman intensity for a two
dimensional array with damping of triplet excitation taken into
account. Note the symmetry of the spectrum, to be contrasted
with Fig. \ref{fig:univ1d}.}
\label{fig:logpeaks}
\end{figure}

\end{document}